\documentclass[useAMS,usenatbib]{mn2e}
\usepackage{graphicx}
\usepackage{hyperref}
\usepackage{amssymb}
\usepackage{amsmath}
\usepackage{color}
\usepackage{natbib}
\usepackage[space]{grffile} 
\usepackage{epstopdf}

\newcommand{\mnras}{MNRAS}
\newcommand{\apj}{ApJ}
\newcommand{\apjs}{ApJS}

\newcommand{\aj}{AJ}
\newcommand{\apjl}{ApJL}
\newcommand{\aap}{A\&A}

\newcommand{\araa}{ARA\&A}

\begin{document}


\title[Cosmic distribution of SMBHs from Pop III.1 Seeds]{The Formation of Supermassive Black Holes from Population III.1 Seeds. I. Cosmic Formation Histories and Clustering Properties}

\author[Banik, Tan \& Monaco]{\parbox{\textwidth}{
Nilanjan Banik\thanks{Email: banik@phys.ufl.edu}$^{1,2,3,4}$, 
Jonathan C. Tan$^{5,6}$, 
Pierluigi Monaco$^{7,8,9}$
}
  \vspace*{4pt} \\ 
$^{1}$Dept. of Physics, University of Florida, Gainesville, FL 32611, USA, \\
$^{2}$Fermi National Accelerator Laboratory, Batavia, IL 60510-0500, USA\\
$^{3}$GRAPPA Institute, Institute for Theoretical Physics Amsterdam\\ 
and Delta Institute for Theoretical Physics, University of Amsterdam, Science Park 904, 1098 XH Amsterdam, The Netherlands, \\
$^{4}$Lorentz Institute, Leiden University, Niels Bohrweg 2,
Leiden, 2333 CA, The Netherlands, \\ 
$^5$Dept. of Space, Earth \& Environment, Chalmers University of Technology, SE-412 93  Gothenburg, Sweden\\
$^6$Dept. of Astronomy, University of Virginia, 530 McCormick Road, Charlottesville, VA 22904-4325, USA\\
$^{7}$Universit\`a di Trieste, Dipartimento di Fisica, Sezione di Astronomia, via Tiepolo 11, 34143 Trieste, Italy, \\
$^{8}$INAF--Osservatorio Astronomico di Trieste, via Tiepolo 11, 34143 Trieste, Italy, \\
$^{9}$INFN, via Valerio 2, I-34127 Trieste, Italy
}

\maketitle

\begin{abstract}
We calculate cosmic distributions in space and time of the formation
sites of the first, ``Pop III.1'' stars, exploring a model in which
these are the progenitors of all supermassive black holes (SMBHs),
seen in the centers of most large galaxies. Pop III.1 stars are
defined to form from primordial composition gas in dark matter
minihalos with $\sim10^6\:M_\odot$ that are isolated from neighboring
astrophysical sources by a given isolation distance,
$d_{\rm{iso}}$. We assume Pop III.1 sources are seeds of SMBHs, based
on protostellar support by dark matter annihilation heating that
allows them to accrete a large fraction of their minihalo gas, i.e.,
$\sim10^5\:M_\odot$.  Exploring $d_{\rm{iso}}$ from
10--$100\:\rm{kpc}$ (proper distances), we predict the redshift
evolution of Pop III.1 source and SMBH remnant number densities. The
local, $z=0$ density of SMBHs constrains $d_{\rm{iso}}\lesssim
100\:\rm{kpc}$ (i.e., $3\:\rm{Mpc}$ comoving distance at
$z\simeq30$). In our simulated ($\sim60\:\rm{Mpc}$)$^3$ comoving
volume, Pop III.1 stars start forming just after $z=40$. Their
formation is largely complete by $z\simeq25$ to 20 for
$d_{\rm{iso}}=100$ to $50\:\rm{kpc}$. We follow source evolution to
$z=10$, by which point most SMBHs reside in halos with
$\gtrsim10^8\:M_\odot$. Over this period, there is relatively limited
merging of SMBHs for these values of $d_{\rm{iso}}$. We also predict
SMBH clustering properties at $z=10$: feedback suppression of
neighboring sources leads to relatively flat angular correlation
functions.

\end{abstract}

\vspace{2cm}
\section{Introduction: The Origin of Supermassive Black Holes}\label{S:intro}

Baryonic collapse appears to lead to two distinct populations of
objects: (1) stars (and associated planets); (2) supermassive black
holes (SMBHs), i.e., with masses $\gtrsim10^5\:M_\odot$. Accretion to
SMBHs powers active galactic nuclei (AGN) and this feedback is thought
to play a crucial role in the evolution of galaxies, e.g., maintaining
high gas temperatures and thus impeding cooling flows and continued
star formation in galaxy clusters. 

In spite of their importance, as discussed below, there is no settled
theory for the formation of SMBHs. Simulations and models of galaxy
formation and evolution typically make ad hoc assumptions about the
creation of these objects. For example, in the Illustris simulation
(Vogelsberger et al. 2014), following methods developed by Sijacki et
al. (2007) and Di Matteo et al. (2008), SMBHs are simply created with
initial masses of $1.4\times10^5\:M_\odot$ in every dark matter halo
that crosses a threshold mass of $7\times10^{10}\:M_\odot$. Barber et
al. (2016) follow a similar method in the EAGLE simulation. In the
semi-analytic models of Somerville et al. (2008), dark matter halos
with $>10^{10}\:M_\odot$ are seeded with black holes with a variety of
initial masses explored from 100 to $10^4\:M_\odot$. Shirakata et
al. (2016) have also explored the effects of the choice of seed black
hole mass in their semi-analytic models of galaxy formation and
evolution. Black holes are created in every ``galaxy,'' i.e., every
halo that is able to undergo atomic cooling. Their models with massive
($10^5\:M_\odot$) seed black holes lead to a black hole mass versus
bulge mass relation that is too high, especially compared to that
observed for lower mass ($\sim 10^9\:M_\odot$) bulges of dwarf
galaxies (Graham \& Scott 2015). Thus Shirakata et al. prefer models
with lower seed masses, e.g., randomly drawn from $10^3$ to
$10^5\:M_\odot$, but other solutions may also be possible, such as
reducing the SMBH occurrence fraction in lower-mass galaxies. For
example, Fontanot et al. (2015) present a model that explains the
break in the black hole versus bulge relation of Scott \& Graham
(2015) as an indirect effect of stellar feedback in the spheroidal
component of galaxies, starting with $10^5\:M_\odot$ SMBH seeds.

The overall goal of this paper is to explore a relatively new physical
mechanism for SMBH formation as the outcome of the evolution of
Population III.1 stars, which are primordial composition (i.e., Pop
III) stars that are the first objects to form in their local regions
of the universe, thus being undisturbed by the feedback from other
astrophysical sources (McKee \& Tan 2008). In particular, since the
formation locations and times of Pop III.1 stars can be predicted by
standard models of cosmological structure formation based on the
growth of halos of cold dark matter (CDM), we aim to predict the
formation histories and clustering properties of SMBHs forming via
this mechanism. This will allow eventual testing of the model against
future observations of high-redshift SMBHs, as well as the properties
of more local SMBH populations.

\subsection{Constraints from the Properties of Local and Distant SMBHs}

From studies of the local universe, SMBHs have masses $\gtrsim
10^5\:M_\odot$ and are found in the centers of most large galaxies
that have spheroidal stellar components (e.g., Kormendy \& Ho 2013;
den Brok et al. 2015; see reviews by Graham 2016 and Reines \&
Comastri 2016). The lowest mass SMBH that has been reported is the
$\sim50,000\:M_\odot$ example in the nucleus of RGG 118 (Baldassare et
al. 2015). However, a number of nearby dwarf galaxies as well as
spiral galaxies with small bulges, e.g., M33, have estimated upper
limits on the presence of a SMBH that are close to
$\sim10^4\:M_\odot$. No nuclear SMBH has yet been detected directly,
e.g., via its optical or X-ray emission, in a galaxy with a stellar
mass $<10^8\:M_\odot$. There are some claimed indirect detections of
SMBHs in ultracompact dwarf galaxies (UCDs) (Seth et al. 2014; Ahn et
al. 2017) from dynamical modeling, but these UCDs are expected to be
the tidally stripped remnants of more massive galaxies, originally
with $\gtrsim 10^9\:M_\odot$ stellar masses.

There have been some claims for the existence of intermediate-mass
black holes (IMBHs), i.e., with masses in the range $\sim100$ to
$\sim10^5\:M_\odot$ that bridge the gap between stellar mass remnants
and SMBHs. The presence of IMBHs within the centers of globular
clusters (GCs) has been reported based on stellar kinematics. However,
in the recent analysis of Baumgardt (2017) in which grids of $N$-body
simulations with and without IMBHs are compared to 50 observed
Galactic GCs, only one system, $\omega$ Cen (NGC~5139), 

is shown to have a clear kinematic signature that may indicate the
presence of an IMBH (of $\sim40,000\:M_\odot$). However, as discussed
by Baumgardt (2017), this is not an unique interpretation, with other
possibilities being the presence of radially anisotropic velocity
dispersion profiles within the cluster (Zocchi, Gieles \&
H\'enault-Brunet 2016). In another individual case, K{\i}z{\i}ltan,
Baumgardt \& Loeb (2017) have reported a
$2200^{+1500}_{-800}\:M_\odot$ black hole in the center of the globular
cluster 47 Tucanae based on the observed kinematics of
pulsars. However, more generally there is no evidence yet for the
expected accretion signatures of IMBHs in globular clusters (e.g.,
Kirsten \& Vlemmings 2012; Wrobel, Miller-Jones \& Middleton
2016). For example, Wrobel et al. (2016) report only upper limits
based on deep cm radio continuum observations of 206 GCs in M81,
although their $3\sigma$ upper limit on the mean black hole mass of
$\sim50,000\:M_\odot$ in the most massive GCs is not that restrictive
and is dependent on the modeling of accretion and radio emission of
the putative IMBHs.

Ultra-luminous X-ray sources (ULXs) away from the nuclei of their host
galaxies and with X-ray luminosities $L_X>10^{39}\:{\rm erg\:s^{-1}}$
(i.e., greater than the Eddington luminosity of a $10\:M_\odot$ black
hole) have been detected and proposed as being evidence for IMBHs. For
example, the source ESO 243-49 HLX1 with $L_X\sim10^{42}\:{\rm
  erg\:s^{-1}}$ and an estimated black hole mass of $M_{\rm
  BH}\sim10^4$--$10^5\:M_\odot$ in a cluster with a stellar mass of
$M_{*}\sim10^5$--$10^6\:M_\odot$ has been claimed by Farrell et
al. (2014). The source M82 X1 with $L_X\sim5\times10^{40}\:{\rm
  erg\:s^{-1}}$ from a $400\:M_\odot$ black hole has been discussed by
Feng et al. (2010) and Pasham et al. (2014). However, NGC~5643 ULX1
with $L_X>10^{40}\:{\rm erg\:s^{-1}}$ has been modeled as a
$30\:M_\odot$ black hole that is undergoing super-Eddington accretion
and/or beaming its emission preferentially in our direction (Pintore
et al. 2016). Overall, there are relatively few clear examples of ULXs
that present unambiguous evidence for IMBHs, with the majority thought
to be explainable as stellar mass black holes in X-ray binaries that
are undergoing active accretion from massive stellar companions
(Zampieri \& Roberts 2009; Feng \& Soria 2011).

Some SMBHs appear to have reached masses $\sim 10^9\:M_\odot$ by
$z\simeq7$ ($t\simeq 800$~Myr after the Big Bang) (e.g., Mortlock et
al. 2011; however, see the factor of $\sim$5 lower revised mass
estimates of Graham et al. 2011; Shankar et al. 2016). However, such
objects are rare: an estimate of the $z\sim6$ quasar luminosity
function finds a number density of observed sources of $\sim
10^{-8}\:{\rm Mpc}^{-3}$ (Willott et al. 2010; see also Treister et
al. 2013).

There seems to be a relative dearth of actively accreting lower-mass
SMBHs at $z\sim6$, based on the flat faint-end slope of the X-ray
luminosity function (XLF) derived from a stacking analysis of the
Chandra Deep Field South (Vito et al. 2016) and the lack of X-ray AGN
in $z\gtrsim6$ Lyman break galaxies (Cowie, Barger \& Hasinger 2012;
Fiore et al. 2012; Treister et al. 2013).

In summary, there appears to be a characteristic minimum mass of SMBHs
of $\sim 10^5\:M_\odot$, with most low-mass galaxies lacking the
presence of any such object, and relatively limited evidence for
IMBHs. Such properties of the SMBH population are a constraint on
theories of their formation. In particular, they may indicate that the
initial seed mass is relatively massive, i.e.,
$\sim10^5\:M_\odot$, and that not all galaxies are seeded with SMBHs.

\subsection{Theoretical Models of SMBH Formation}

SMBH formation scenarios have been discussed for many years (e.g.,
Rees 1978). One popular model is ``direct collapse'' of massive,
primordial composition gas clouds, which is thought to require strong
UV (Lyman-Werner) radiation fields to dissociate $\rm H_2$ molecules
and thus prevent cooling to $\sim200$~K and fragmentation to
$\sim100\:M_\odot$ mass scales, but also requires dark matter halo
virial temperatures $\gtrsim8,000\:$K (i.e., masses $\gtrsim
10^8\:M_\odot$) (e.g., Haehnelt \& Rees 1993; Bromm \& Loeb 2003;
Begelman et al. 2006; Dijkstra et al. 2006; Ferrara \& Loeb 2013;
Dijkstra et al. 2014; Chon et al. 2016).  The high accretion rates
  that occur in the centers of these halos may allow the formation of
  supermassive stars, which then collapse to form SMBHs (e.g.,
  Inayoshi et al. 2013; Umeda et al. 2016).  The study of Chon et
al. (2016) examined a $(20 h^{-1}\:{\rm Mpc})^3$ volume to search for
dark matter halos meeting these criteria, finding about 50 candidates
that form at $z\sim10$ to 20. However, only two of these were seen to
undergo collapse, with the others mostly disrupted by mergers, tidal
disruptions and/or ram pressure stripping with neighboring
halos. While the number density of successful direct collapse events
in their model, i.e., $\sim 10^{-4}\:{\rm Mpc}^{-3}$, is greater than
the observed number density of high-$z$ SMBHs, it is much smaller than
the total comoving number density of SMBHs observed at $z=0$
($\sim10^{-3}$--$10^{-2}\:{\rm Mpc}^{-3}$; see \S\ref{S:nevol}), so it
may be difficult for this mechanism to form all SMBHs. One possibility
that could boost the number density of direct collapse events,
discussed by Chon et al., is that their adopted critical UV flux to
prevent $\rm H_2$ formation is too conservative. However, as also
discussed by Chon et al., the modeling of direct collapse halos to
make accurate predictions of event rates is very challenging, since it
requires making a number of uncertain assumptions about the star
formation in the early universe that sets the UV feedback environments
necessary for this model.

An alternative model that may create the conditions for direct
collapse has been proposed by Mayer et al. (2010; 2015) involving
rapid infall of gas driven by mergers of gas-rich massive galaxies
occurring at $z\sim4$--$10$. This scenario does not require primordial
composition gas, with fragmentation and star formation of the gas
suppressed by gravity-driven turbulence and torques. Up to
$\sim10^9\:M_\odot$ of gas is proposed to be able to accumulate on
sub-parsec scales, leading to ultra-massive black hole seeds of
$\sim10^8\:M_\odot$. However, as discussed by Mayer et al. (2015),
caveats of this model include that the supporting simulation results
are based on binary collisions of gas-rich galaxies rather than being
self-consistently extracted from cosmological simulations. Estimates
of the frequency of black hole formation via this scenario are quite
uncertain, although potentially high enough to explain the observed
high $z$ quasar population.

Yet another model involves high ($\gtrsim200\:{\rm km\:s^{-1}}$)
velocity collisions of protogalaxies that create hot, dense gas that
leads to collisional dissociation of $\rm H_2$ molecules (Inayoshi \&
Omukai 2012; Inayoshi et al. 2015). However, the comoving number
density of black holes formed by such a mechanism is estimated to be
only $\sim 10^{-9}\:{\rm Mpc}^{-3}$ by $z\sim10$, which, while it may
be enough to help explain observed high $z$ quasars, is too small to
explain all SMBHs.

Another scenario for IMBH or SMBH formation involves a very massive
stellar seed forming in the center of a dense stellar cluster by a
process of runaway mergers (G\"urkan et al. 2004; Portegies Zwart et
al. 2004; Freitag et al. 2006). However, the required central stellar
densities are extremely high (never yet observed in any young cluster)
and also stellar wind mass loss may make growth of the central very
massive star quite inefficient (Vink 2008), unless the metallicities
are very low (Devecchi et al. 2012). Again, predictions of such models
for the cosmic formation rates of SMBHs are limited since it is
difficult to predict when and how the necessary very dense star
clusters are formed.

Finally a class of models involve SMBHs forming from the remnants of
Pop III stars, i.e., those forming from essentially metal-free gas
with compositions set by big bang nucleosynthesis (see, e.g., Bromm
2013 for a review). McKee \& Tan (2008) distinguished two classes of
Pop III stars. Pop III.1 are those that form in isolation, i.e.,
without suffering significant influence from any other astrophysical
source (i.e., other stars or SMBHs). Molecular hydrogen cooling leads
to $\sim 200$~K temperatures in the centers of minihalos and first
unstable fragment scales of $\sim100\:M_\odot$ (Bromm et al. 2002;
Abel et al. 2002).  Pop III.2 stars still have primordial composition,
but are influenced by external astrophysical sources, with the most
important effects expected to be due to radiation feedback from
ionizing or dissociating radiation (e.g., Whalen et al. 2008). One
effect is to photoevaporate the gas from the minihalos, thus delaying
star formation until the halos are more massive. The masses of Pop
III.2 stars are thought to be potentially smaller than those of Pop
III.1 stars due to enhanced electron fractions in gas that has
suffered greater degrees of shock heating and/or irradiation that then
promotes greater rates of $\rm H_2$ and HD formation and thus more
efficient cooling and fragmentation (e.g., Greif \& Bromm 2006).

However, although the initial unstable baryonic mass scale is commonly
thought to be $\sim 100\:M_\odot$ in Pop III.1 halos set by the
microphysics of $\rm H_2$ cooling, the ultimate masses of the stars
that form are quite uncertain. The initial $\sim100\:M_\odot$ unstable
``core'' is typically located at the center of the dark matter
minihalo and surrounded by an envelope of $\sim10^5\:M_\odot$ of gas
that is bound to the halo. Tan \& McKee (2004) and McKee \& Tan (2008)
presented semi-analytic estimates for final accreted masses of Pop
III.1 protostars of $\sim 140\:M_\odot$ set by disk photoevaporation
feedback. Using improved protostellar evolution models, Tanaka, Tan \&
Zhang (2017) have revised these estimates to $\sim
50\:M_\odot$. Hosokawa et al. (2011) found similar results using
radiation-hydrodynamic simulations. Tan, Smith \& O'Shea (2010)
applied the MT08 model to accretion conditions in 12 minihalos from
the simulations of O'Shea et al. (2006), i.e., that have a variety of
accretion rates, and estimated an initial mass function that peaked at
$\sim 100\:M_\odot$ with a tail extending to $\sim 10^3\:M_\odot$ from
those sources that have the highest accretion rates. Hirano et
al. (2014) and Susa et al. (2014) presented radiation-hydrodynamic
simulations of populations of $\sim100$ Pop III.1 stars, deriving
initial mass functions peaking from $\sim 10$--$100\:M_\odot$, with a
tail out to $\sim 10^3\:M_\odot$. Their formation redshifts extended
from $z\sim35$ down to $z\sim 10$.

While the above estimates for Pop III.1 masses are certainly top heavy
compared to present-day star formation, they are still relatively low
compared to the masses of SMBHs. To reach $\sim10^9\:M_\odot$ in a few
hundred Myr to explain the observed high-$z$ quasars would require
near continuous maximal (Eddington-limited) accretion. Such accretion
seems unlikely given that massive Pop III.1 stars would disrupt the
gas in their natal environments by radiative and mechanical feedback
(e.g., O'Shea et al. 2005; Johnson \& Bromm 2007; Milosavljev\'ic et
al. 2009).

Several authors have invoked the effect of coherent relative streaming
velocities between dark matter and gas (Tseliakhovich \& Hirata 2010),
which then leads to more massive (on average by about a factor of 3;
Greif et al. 2011b; see also Fialkov et al. 2012 and Schauer et
al. 2017) minihalos being the sites of Pop III.1 star formation, as a
mechanism that may lead to conditions of SMBH formation. The idea is
that in the rarer cases where a minihalo is forming in a region where
the streaming velocities are significantly ($\gtrsim 2\times$) larger
than average, then the minihalo mass at time of first star formation
is also larger, perhaps $\sim10^7\:M_\odot$ with a virial temperature
$\sim8,000\:$K (Tanaka \& Li 2014). Collapse in such a halo would
proceed at a relatively high accretion rate that can lead to
protostellar swelling (Hosokawa et al. 2016) that reduces ionizing
feedback. Simulations of such a model to form a 34,000$\:M_\odot$
protostar have been presented by Hirano et al. (2017). This mechanism
is potentially attractive, especially since it is relatively simple in
being able to predict the formation locations of the sites of SMBH
formation (Tanaka \& Li 2014). However, whether or not this mechanism
can produce sufficient numbers of SMBHs and whether a minimum
characteristic mass of $\sim10^5\:M_\odot$ can naturally be produced,
rather than a continuous distribution with large numbers of IMBHs,
remains to be determined.

In the next subsection we discuss how the outcome of Pop III.1 star
formation may be altered under the influence of the energy input from
Weakly Interacting Massive Particle (WIMP) dark matter
self-annihilation. This mechanism may provide a route for the
formation of supermassive, i.e., $\sim10^5\:M_\odot$, Pop III.1 stars,
which would then collapse to form SMBHs. Furthermore, this mechanism,
which requires special conditions of the co-location of the protostar
with the central density cusp of the dark matter halo, is only
expected to be possible in Pop III.1 sources. This opens up the
possibility of a ``bifurcation'' in the collapse outcome, i.e.,
$\sim10^5\:M_\odot$ SMBHs from Pop III.1 sources and
$\lesssim100\:M_\odot$ stellar populations from all other
sources. Another attractive feature of this scenario for SMBH
formation is its relative simplicity, with relatively few free
parameters. It is thus amenable to incorporation into semi-analytic
models of structure and galaxy formation to make testable predictions.

\subsection{Population III.1 Dark Matter Annihilation Powered Protostars as SMBH Progenitors}

One potential mechanism that may allow supermassive Pop III.1 stars to
form is energy injection inside the protostar (i.e., during the
accretion phase) by WIMP dark matter self-annihilation (Spolyar et
al. 2008; Natarajan, Tan \& O'Shea 2009, hereafter NTO09). As
discussed below, this energy injection can be sufficient to support
the protostar in a very large, swollen state, which gives it a
relatively cool photospheric temperature and thus relatively weak
ionizing feedback. If this state of weak feedback can be maintained as
the protostar accretes the baryonic content of its minihalo, i.e.,
$\sim10^5\:M_\odot$, then this provides a pathway to create a
supermassive star, which would be expected to soon collapse to a SMBH.

NTO09 estimated that for the early phases of protostellar evolution to
be significantly affected by WIMP annihilation heating, the WIMP mass
needs to be $m_\chi \lesssim$ several~$\times100$~GeV, based on the
size of the initial protostellar core in which WIMP heating dominated
over baryonic cooling. Such WIMP masses are consistent with
constraints on $m_\chi$ that are based on constraints on the WIMP
annihilation cross section along with the requirement that the actual
cross section should equal the thermal relic value of $\sim3\times
10^{-26}\:{\rm cm^3\:s^{-1}}$ (i.e., that necessary for all, or most,
dark matter to be composed of WIMPs). For example, from particle
production in colliders, Khachatryan et al. (2016) find $m_\chi\gtrsim
6$ to 30~GeV depending on whether the process is mediated via vector
or axial-vector couplings of Dirac fermion dark matter. From indirect
searches via Fermi-LAT observations of expected gamma ray emission
from 15 Milky Way dwarf spheroidal satellite galaxies, including
assumptions for modeling their dark matter density structures,
Ackermann et al. (2015) report $m_\chi\gtrsim 100$~GeV for WIMPs
annihilating via quark and $\tau$-lepton channels. Such results
suggest there may be only a relatively narrow range of WIMP masses
that are viable for this scenario of dark matter powered Pop III.1
protostars.

Also, direct detection experiments that constrain (spin dependent and
independent) WIMP-nucleon elastic scattering cross sections have yet
to detect signatures of WIMP dark matter (e.g., Agnese et al. 2014;
Akerib et al. 2016). While these results do not provide a firm
constraint on the value of $m_\chi$, they do have implications for the
ability Pop III.1 protostars to capture WIMPs of a given mass via such
scattering interactions.

Spolyar et al. (2009) followed the growth of Pop III.1 protostars
including the effects of annihilation heating from their initial and
captured dark matter. We note that such dark matter powered protostars
are objects that collapse to densities quite similar to those of
normal protostars: e.g., for the initial model considered by Spolyar
et al. (2009), which has $3\:M_\odot$ and soon achieves a radius of
$\sim 3\times 10^{13}\:{\rm cm}$, i.e., $\sim 2$~AU, the central
densities are $n_{\rm H,c}\sim6\times10^{17}\:{\rm cm^{-3}}$ and the
mean densities $\bar{n}_{\rm H}\sim3\times 10^{16}\:{\rm cm^{-3}}$. In
their cannonical case without captured WIMPs, collapse of the
protostar to the main sequence was delayed until about
$800\:M_\odot$. They also considered a ``minimal capture'' case with
background dark matter density of $\rho_\chi=1.42\times10^{10}\:{\rm
  GeV\:cm}^{-3}$ and $\sigma_{\rm sc}=10^{-39}\:{\rm cm}^2$ (for
spin-dependent scattering relevant to H) that results in about half
the luminosity being from annihilation of captured WIMPs (with
$m_\chi=100$~GeV) and the other half from nuclear fusion at the time
the protstar joins the main sequence. The results from Akerib et
al. (2016) imply $\sigma_{\rm sc}\lesssim 5\times 10^{-39}\:{\rm
  cm}^2$. However, more recent studies have lowered this to
$\sigma_{\rm sc}\lesssim 8\times 10^{-40}\:{\rm cm}^2$ (Akerib et
al. 2017) and $\sigma_{\rm sc}\lesssim 5\times 10^{-41}\:{\rm cm}^2$
for a WIMP mass of 100~GeV, which thus call into question the validity
of the minimal capture model and require consideration of other
capture models.

Freese et al. (2010) and Rindler-Daller et al. (2015) have presented
models of protostellar evolution of dark matter annihilation powered
protostars that continue to accrete to much higher masses. In the
study of Rindler-Daller et al., starting with initial protostellar
masses from 2 to $5\:M_\odot$, the protostars are followed to $\gtrsim
10^5\:M_\odot$ for cases with accretion rates of $10^{-3}$ and
$10^{-1}\:M_\odot\:{\rm yr}^{-1}$ and WIMP masses of $m_\chi=10, 100,
1000\:$GeV. While feedback effects, i.e., ionization, are not
considered that may limit the accretion rate, the protostars tend to
remain relatively large and thus cool, especially for the $m_\chi=10$
and 100~GeV cases. 

The requirements for forming supermassive Pop III.1 protostars, which
then collapse to form SMBHs, can be summarized as follows. The
ionizing luminosity needs to remain low compared to that of a
protostar of the equivalent mass on the zero age main sequence (ZAMS),
else disk photoevaporation (McKee \& Tan 2008; Hosokawa et al. 2011;
Tanaka et al. 2017) will shut off the accretion flow. There appear to
be a range of models for $m_\chi\lesssim1$~TeV in which this occurs, which
could allow continued accretion of the baryons within the minihalo, in
principle up to the entire baryonic content $\sim10^5\:M_\odot$ 
  (Rindler-Daller et al. 2015). Indeed, efficient accretion of the gas
  of the minihalo to the central protostar is a requirement and
  feature of this model. For this to occur, additional requirements
are that angular momentum can be lost from the gas and that
fragmentation does not occur to create gravitational fluctuations that
scatter and thus dilute the central dark matter density or divert a
significant fraction of the baryonic mass flux of the collapsing
minihalo to a binary companion or multiple companions. Especially
binary formation that leads to displacement of the primary star from
the ``central region'' of the dark matter halo could then shut off
continued capture of dark matter to the protostar, which may be
necessary to achieve the highest masses.

The question of angular momentum transport is one that has already
been studied in the context of traditional models and simulations of
Pop III.1 star formation. Abel, Bryan \& Norman (2002) showed that the
specific angular momentum of the gas as a function of radius in the
minihalo maintained a fairly constant, sub-Keplerian level during the
early phases of infall. Significant angular momentum transport was
achieved by trans-sonic turbulent motions, driven by the gravitational
contraction. A key feature of this infall is that it occurs relatively
slowly, mediated by the relatively weak rates of $\rm H_2$
ro-vibrational line cooling, so the gas in the minihalo is in
approximate pressure and virial equilibrium. This allows sufficient
time for angular momentum transport within the gas cloud. 

At later stages of collapse, after the first Pop III.1 protostar has
formed, WIMP annihilation heating is only expected to be significant
in the protostar if these objects are co-located near the peaks of
their natal dark matter minihalos and the dark matter density stays
sufficiently high in these zones.

Spolyar et al. (2009) have presented a case in which effective WIMP
annihilation heating alters the evolution of the protostar for a dark
matter density of $\rho_\chi \gtrsim 10^{10}\:{\rm GeV}\:{\rm
  cm}^{-3}$. For the three example minihalos considered by NTO09, the
initial radii defining such a central region are 72, 45 and 89~AU.
Stacy et al. (2014) carried out a simulation that followed Pop III.1
star formation including an ``active'' dark matter halo that has its
central density reduced by interactions with a clumpy accretion disk
around the primary protostar. The radius of the zone that has
$\rho_\chi \gtrsim 10^{10}\:{\rm GeV}\:{\rm cm}^{-3}$ was still
$\sim40$~AU at the end of their simulation, 5000~yr after first
protostar formation, and the primary protostar was located within this
zone.

In terms of fragmentation of the gas, this was seen to be relatively
limited during the initial phases of collapse of minihalos in the
cosmological simulations of Turk, Abel \& O'Shea (2009): about 80\% of
their minihalos appear to collapse to a single protostellar ``core,''
i.e., the self-gravitating gas in which a single rotationally
supported disk will form. However, a number of authors have claimed
that later fragmentation of the primary protostar's accretion disk may
lead to formation of multiple lower mass stars leading to either
formation of a close binary or even a cluster of low-mass Pop III
stars. Clark et al. (2011) followed the evolution to about 110~yr
after first protostar formation, including effects of protostellar
heating on the disk. By this time the protostar had accreted almost
0.6~$M_\odot$. It still resided near the center of its accretion disk,
which, being gravitationally unstable, had also formed three lower
mass ($\lesssim0.15\:M_\odot$) protostars. Greif et al. (2011a)
followed the collapse and fragmentation of five different minihalos to
about 1000~yr after first protostar formation, by which time masses of
several solar masses had been achieved. They typically observed
several tens of protostars forming by fragmentation in the main
accretion disk, although describe that most are likely susceptible to
merging with the primary protostar (a process they could not follow in
their simulations).

Smith et al. (2012) carried out similar simulations, but now including
the effects of WIMP annihilation on the chemistry and thermodynamics
of the collapse. They found much reduced fragmentation: in one case
only a single, primary protostar formed; in another, just one
secondary protostar. The primary protostars reached $\gtrsim
10\:M_\odot$ and remained in the central regions of their host
minihalos, which would imply there could be an important effect on
subsequent protostellar evolution due to WIMP annihilation
heating. Stacy et al. (2014) also carried out such simulations, but
now with no protostellar heating feedback. They formed a primary
protostar that reached $8\:M_\odot$ after 5000~yr along with several
lower-mass companions. This primary protostar was still located in a
zone with dark matter density $\rho_\chi \gtrsim 10^{10}\:{\rm
  GeV}\:{\rm cm}^{-3}$.

Another point that should be noted is that the propensity of
protostellar accretion disks to fragment will also depend on the
magnetic field strength in the disk and none of the above
fragmentation studies have included $B$-fields. Dynamo amplification
of weak seed fields that arise via the Biermann battery mechanism may
occur in turbulent protostellar disks: see, e.g., Tan \& Blackman
(2004); Schleicher et al. (2010); Schober et al. (2012); Latif \&
Schleicher (2016). These studies all predict that dynamically
important, near equipartition $B$-fields will arise in Pop III
protostellar disks. From numerical simulations of local star formation
it is well known that such $B$-fields are important for enhancing
transport of angular momentum during collapse and also for generally
acting to suppress fragmentation compared to the unmagnetized case
(e.g., Price \& Bate 2007; Hennebelle et al. 2011). Magnetic fields
would similarly be expected to reduce the density fluctuations in the
accretion disks and the size of the disks, which would then reduce the
amount of gravitational interaction that is seen to dilute the dark
matter density cusp in the simulation of Stacy et al. (2014). If early
fragmentation is suppressed then this may allow the primary protostar
to achieve a significant mass and luminosity so that its radiative
feedback then later becomes the dominant means of limiting
fragmentation.

In summary, whether or not a single dominant, centrally-located
protostar is the typical outcome of collapse of Pop III.1 minihalos is
still uncertain. Such outcomes are seen in some pure hydrodynamic
simulations of collapse, especially when the effects of dark matter
annihilation heating are included. If magnetic fields can be amplified
to near equipartitition by an accretion disk dynamo, then this outcome
is expected to be even more likely to occur.

On the other hand, Pop III.2 stars, if formed in a minihalo that
undergoes very significant early stage fragmentation to multiple
``cores,'' are not generally expected to be co-located with the dark
matter density peak. Co-location will also not occur for stars forming
in more massive halos, where first atomic cooling and then $\rm H_2$
or metal or dust cooling allows formation of a large-scale
rotationally-supported thin disk, which then fragments to form a more
normal stellar population, i.e., the early stages of a galactic disk.

We thus regard formation of dark matter powered Pop III.1 stars as a
potentially attractive mechanism to explain the origin of SMBHs,
possibly all SMBHs.
If the protostar is in a large, swollen state, as is generally
  expected if WIMP annihilation heating is important, and is thus able
  to accrete a significant fraction of the initial baryonic content of
  the Pop III.1 minihalo, i.e., $\gtrsim 10^5\:M_\odot$,

then this is likely to lead to SMBH formation via an intermediate
stage of supermassive star formation. Collapse to a SMBH may be
induced by the star becoming unstable with respect to the general
relativistic radial instability (GRRI). For nonrotating main sequence
stars this is expected to occur at a mass of
$\simeq5\times10^4\:M_\odot$ (Chandrasekhar 1964), while in the case
of maximal uniform rotation this is raised to $\sim10^6\:M_\odot$
(Baumgarte \& Shapiro 1999).

SMBH formation from supermassive Pop III.1 stars that efficiently
accrete the baryons from their minihalos is thus a mechanism
that can help explain the apparent absence or dearth of SMBHs with
masses $\lesssim10^5\:M_\odot$. This lower limit to the masses of
SMBHs is not easily explained in most other formation models.

\subsection{Goals and Outline of this Paper}

Our goals in this paper are to make predictions for cosmological
populations of SMBHs that form via Pop III.1 protostars supported by
dark matter annihilation heating. We note that a broad consensus on
the validity of this mechanism as an outcome of Pop III.1 star
formation has not yet been reached (see, e.g., Clark et al. 2011;
Greif et al. 2011a; Smith et al. 2012; Stacy et al. 2014). However,
here we will assume the validity of this model in order to follow its
consequences and predictions. Such predictions, especially the
formation history of Pop III.1 stars, the overall number densities of
these stars and their proposed SMBH remnants, and their clustering
properties, are necessary as a first step to eventually connect to
observations of SMBH populations at high and low redshift and thus
test this theoretical model of SMBH formation.

The conditions needed to be a Pop III.1 protostar, i.e., for being
``undisturbed'' by other astrophysical sources, so that the protostar
is co-located with the dark matter cusp to enable effective WIMP
annihilation heating, are uncertain. This is because the radiative
influence on a halo from a neighboring source and its effect on
subsequent star formation is a very complicated problem (e.g., Whalen
\& Norman 2006), which also depends on the nature of the sources of
feedback. For simplicity we will therefore first parameterize the
required ``isolation distance,'' $d_{\rm iso}$, that is needed for a
given halo to be a Pop III.1 source and consider a range of
values.

The outline of the paper is as follows. In \S\ref{S:methods} we
present our methods for simulating structure formation and identifying
Pop III.1 minihalos. In \S\ref{S:results} we present our main results,
i.e., the evolution of the number densities of Pop III.1 stars and
SMBH remnants (\S\ref{S:nevol}), the sensitivity of the results to
cosmic variance and the cosmological parameter $\sigma_8$
(\S\ref{S:variance}), the mass function of SMBH host halos in the post
formation phase at $z=10$ and 15 (\S\ref{S:massfn}), synthetic sky
maps of the sources (\S\ref{S:maps}), and evaluation of the angular
correlation function of the predicted SMBH populations in
(\S\ref{S:angular}).

We discuss the implications of our results and draw
conclusions in \S\ref{S:disc}.

\section{Methods}\label{S:methods}

We utlilize PINOCCHIO (PINpointing Orbit Crossing Collapsed
HIerarchical Objects), which is a code based on Lagrangian
Perturbation Theory (LPT) (Moutarde et al. 1991; Buchert \& Ehlers
1993; Catelan 1995)

for the fast generation of catalogs of dark matter halos in
cosmological volumes.   LPT (see review by Monaco 2016) is
  a perturbative approach to the evolution of overdensities in a
  matter-dominated Universe. It is based on the Lagrangian description
  of fluid dynamics, and its validity is mainly limited to laminar
  flows, where the orbits of mass elements do not cross. As such, this
  is ideal to describe the early universe, characterised by a limited
  degree of non-linearity. Starting from a realization of a Gaussian
density field in a box sampled by $N^3$ particles,

using an ellipsoidal collapse model, PINOCCHIO computes the time at
which each particle is expected to suffer gravitational collapse
(i.e., ``orbit crossing,'' when the map from initial, Lagrangian, to
final, Eulerian positions becomes multivalued), then collects the
collapsed particles into halos with an algorithm that mimics their
hierarchical clustering. The result is a catalog of dark matter halos
with known mass, position, velocity and merger history.

The code was introduced in its original form by 
Monaco, Theuns \& Taffoni (2002),

where it was demonstrated that it can accurately reproduce
``Lagrangian'' quantities like halo masses and merger histories. This
was later confirmed by other groups (see the review in Monaco et
al. 2013).
  
A massively parallel version was presented by Monaco et al. (2013)

and extended and tested by Munari et al. (2016).

We have compared the mass function of halos produced by our PINOCCHIO
simulations with the analytic fit to the N-body simulations of Reed et
al (2007), finding generally very good agreement, but with two
caveats. First, at $z=30$, while the number of minihalos of mass
$\sim10^6\:M_\odot$ predicted by PINOCCHIO agrees to within $\sim20\%$
with the Reed et al. results, there is a modest deficit of higher-mass
halos by about a factor of 2, potentially caused by finite volume
effects (see Reed et al. 2007; Monaco 2016). For the purposes of this
paper, we are primarily concerned with predicting the emergence of
minihalos and so consider the results of PINOCCHIO to be sufficiently
accurate for such purposes. Second, at $z=10$ there is a mild
overestimate of $\gtrsim 10^8\:M_\odot$ halos by about a factor of
1.5, due to the fact that the PINOCCHIO mass function has been
calibrated on the numerical fit of Watson et al. (2014), which gives
more massive halos in the high mass tail. These discrepancies give a
measure of the uncertainties in our numerical halo mass function
estimates.

For clustering properties, Munari et al. (2016) showed how PINOCCHIO's
prediction of the clustering of halos improves when higher orders of
LPT are used.  As a result, clustering in $k$-space is well
reproduced, to within a few per cent, up to a wavenumber of at least
$k=0.3\ h$ Mpc$^{-1} = 0.203\ h_{0.68}$ Mpc$^{-1}$, where
$h_{0.68}\equiv h/0.6774 = 1$ is the normalized Hubble parameter and
will be used in lieu of $h$ henceforth. A degradation of quality is
seen at $z<0.5$, where the density field becomes significantly
non-linear.  Clustering in configuration space is very well reproduced
on comoving scales larger than $\sim(10 - 20) h^{-1}$ Mpc =
$(14.76-29.52) h_{0.68}^{-1}$ Mpc.

In this paper we use the latest code version with 2LPT (2nd order)
displacements, that give a very good reproduction of clustering while
keeping memory requirements to $\sim150$ bytes per particle, thus
allowing running of large boxes (3LPT would require nearly twice as
much memory). PINOCCHIO is well-suited to the study of the formation
of first stars and SMBHs from high-$z$, relatively low-mass halos
spanning large cosmological volumes.

Adopting a standard Planck cosmology: $\Omega_{\rm{m}}=0.3089$,
$\Omega_{\Lambda}=0.6911$, $n_{\rm{s}}=0.9667$, $\sigma_{8}=0.8159$,
$\Omega_{b}h^{2}=0.02230 = \Omega_{b}h_{0.68}^{2}=0.0102$, $w_{0}=-1$,
$w_{1}=0$ (Planck collaboration 2015), we simulate a 40.96 $h^{-1}=
60.47 h_{0.68}^{-1}$ Mpc comoving cubical box sampled with $4096^3$
particles, thus reaching a particle mass of $1.2\times 10^{5} \, {\rm
  M}_\odot$. This allows us to sample a $10^6\ M_\odot$ halo with
$\sim$10 particles.

We first run the simulation down to redshift of 15. This required 10
Pb of RAM and took less than an hour on 1376 cores of the
GALILEO@CINECA machine, most of the time being spent in writing 211
outputs from $z=40$ to $z=10$ in redshift steps of $\Delta z=0.1$. We
then continued the simulation down to $z=10$, outputing in steps of
$\Delta z =1$. This is the largest PINOCCHIO run ever presented in a
paper.

From the simulation outputs, we identify halos that form Pop III.1
stars by looking for minihalos with masses just crossing a threshold
of $10^{6}\:M_{\odot}$, which are also isolated from any other
existing minihalos (i.e., that may host Pop III.1, Pop III.2 or Pop II
sources) by a proper distance of $d_{\rm iso}$. This assumption of a
constant threshold halo mass of $10^6\:M_\odot$ is motivated first by
its simplicity. The masses of the dark matter halos of the $\sim$100
Pop III stars studied in the simulations of Hirano et al. (2014), have
a fairly narrow mass distribution around $\sim 3\times
10^5\:M_\odot$. However, the effect of coherent relative streaming
velocities between dark matter and gas (Tseliakhovich \& Hirata 2010)
have been shown to delay Pop III star formation, i.e., increasing the
required halo mass by about a factor of three (Greif et al. 2011b),
which was not allowed for in the study of Hirano et al. (2014). Given
these considerations, we regard the choice of a constant threshold
mass of $10^6\:M_\odot$ as a reasonable first approximation. However,
we have also explored the effects of varying this choice of threshold
mass up to values of $4\times10^{6}\:M_\odot$, which, as we describe
below, only have very minor effects on our main results.

On the other hand, the main parameter that we explore in this model of
SMBH formation from Pop III.1 sources is the isolation distance,
$d_{\rm iso}$, with values of $d_{\rm iso} = 10, 20, 30, 50, 100,
300$~kpc being considered.

Once halos have been tagged as being Pop III.1 sources we then track
their subsequent evolution. First, for the next period of time,
$t_{*f}$, they are considered to be ``Pop III.1 Stars'', which
includes the protostellar accretion phase and any additional period of
stellar evolution. We examine specific choices of $t_{*f}=10$, 30 and
100~Myr. For $10^5\:M_\odot$ stars that have negligible post accretion
lifetimes, i.e., if accreting right up to the point of GRRI, this
corresponds to average accretion rates in the range $10^{-3}$ to
$10^{-2}\:M_\odot\:{\rm yr}^{-1}$, which are typical values expected
for such sources (e.g., Tan \& McKee 2004; Tan et al. 2010; see also
the pre-feedback accretion rates in the simulations of Hirano et
al. 2014).

After $t_{*f}$, Pop III.1 stars are assumed to collapse into SMBH
remnants. We note that while the value of $t_{*f}$ is quite uncertain,
it does not affect the eventual properties of the SMBH remnants. The
halos containing SMBHs are tracked down to $z=10$. These halos grow in
mass by both accretion of dark matter particles (i.e., sub-minihalos)
and by mergers with already identified minihalos and larger
halos. During a merger of two halos, the more massive halo retains its
identity and typically the SMBH will be occupying the more massive of
the two merging halos. Occasionally the SMBH and its host halo merge
with a more massive halo, in which case the presence of the SMBH is
transferred to this new halo. Sometimes two merging halos will each
already host a SMBH: this situation is expected to lead to SMBH-SMBH
binary in the center of the new halo and thus potentially a merger of
the two black holes.

In this paper we focus on SMBH locations, number densities and host
halo properties, and defer modeling of SMBH growth due to gas
accretion (i.e., active galactic nuclei) to a future paper.  While we
note when SMBH mergers are expected to occur, we also defer analysis
of these merger properties and potential gravitational wave signatures
to a future study.

\section{Results}\label{S:results}

We applied the algorithm described in \S\ref{S:methods} to identify
Pop III.1 minihalos and follow their assumed SMBH remnants down to
$z=10$ for the fiducial simulation volume and several other test
volumes. The main results for SMBH formation histories, halo
properties and clustering properties are described in this section.

\subsection{Cosmic Evolution of the Number Density of Pop III.1 Stars and SMBHs}\label{S:nevol}

\begin{figure*}
\centering
\includegraphics[scale=1.0]{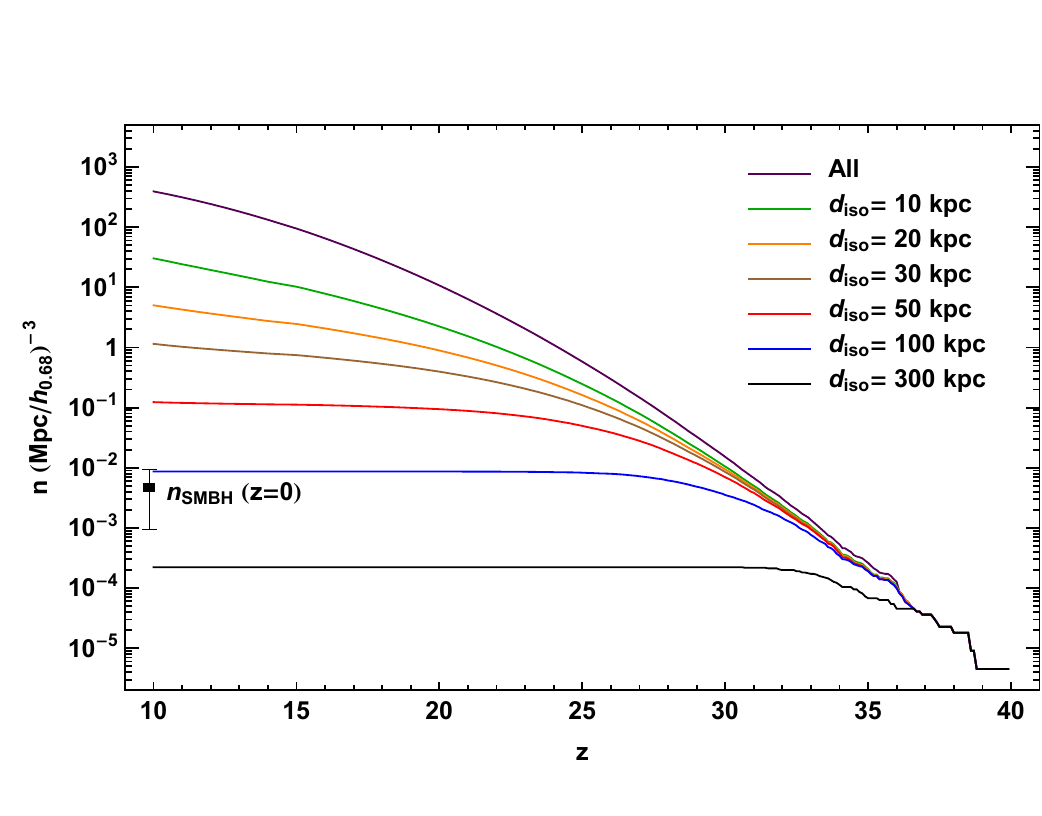}
\caption{
Evolution of comoving total number density of Pop III.1 stars and
their SMBH remnants for different values of $d_{\rm iso}$ ranging from
10 to 300~kpc (proper distances). The number density of all dark
matter halos with $M>10^6\:M_\odot$ is also shown for reference. The
data point $n_{\rm SMBH}(z=0)$ drawn schematically on the left side of
the figure shows an estimate for the present day number density of
SMBHs assuming one SMBH is present in all galaxies with $L>L_{\rm
  min}$. The solid square shows this estimate for $L_{\rm
  min}=0.33L_{*}$, while the lower and upper bounds assume $L_{\rm
  min}=0.1L_{*}$ and $L_{*}$, respectively. }
\label{fig:nevol}
\end{figure*}

Figure~\ref{fig:nevol} shows the redshift evolution of the comoving
total number density, $n$, of Pop III.1 stars and remnants (i.e.,
assumed in this model to be SMBHs) for different values of $d_{\rm
  iso}$ ranging from 10 to 300~kpc (proper distances). As we will see
below, for the assumption of 10 to 100~Myr lifetimes of Pop III.1
stars, these totals soon become dominated by the SMBH remnants. Within
this simulated (60.47 $h_{0.68}^{-1}$~Mpc)$^3$ comoving volume, Pop
III.1 stars start forming just after $z=40$. For large values of
$d_{\rm iso}$, the number of new Pop III.1 sources that are able to
form decreases more quickly and $n$ asymptotes to a constant
value. For example, for $d_{\rm iso}=100$~kpc Pop III.1 stars have
largely ceased to form by $z\sim 25$, while for $d_{\rm iso}=10$~kpc
they continue to form still at $z=10$.

Comparison of the number density of sources formed in these models
with the present day ($z=0$) number density of SMBHs, $n_{\rm SMBH}$,
constrains $d_{\rm iso}$. We estimate $n_{\rm SMBH}(z=0)\sim
0.015\:({\rm Mpc/h})^{-3} = 4.6\times10^{-3}\:({\rm Mpc/h_{0.68}})^{-3}$ by
assuming that all galaxies with $L_{\rm min}>0.33L_*$ host SMBHs and
integrating over the local galaxy luminosity function assumed to be a
Schecter function of the form
\begin{equation}
\phi(L)=\left(\frac{\phi^*}{L^*}\right)\left(\frac{L}{L^*}\right)^{\alpha}e^{-(L/L^*)}
\end{equation}
where $\phi^{*} = 1.6 \times 10^{-2}~ \rm{(Mpc/h)^{-3}}= 4.9\times
10^{-3}~ \rm{(Mpc/h_{0.68})^{-3}}$ is the normalization density and
$L^*$ is the characteristic luminosity corresponding to
$M_{\rm{B}}=-19.7 + 5\log h= -20.55$ (e.g., Norberg et al. 2002) and
$\alpha\simeq1.2$ is the power law slope at low $L$. This value is shown on the
left edge of Figure~\ref{fig:nevol}, with the error bar resulting from
assuming $L_{\rm min}=0.1$ to 1~$L_*$, i.e., a range from $9.3\times
10^{-3} \rm{(Mpc/h_{0.68})^{-3}}$ to $9.3 \times 10^{-4}
\rm{(Mpc/h_{0.68})^{-3}}$. For comparison, integrating the SMBH mass
function from Vika et al. (2009), we estimate the number density of
SMBHs to be $8.79\times10^{-3}\:\rm{(Mpc/h_{0.68})^{-3}}$, which is
consistent with our more simplistic estimate.

\begin{figure*}
\centering
\includegraphics[scale=0.8]{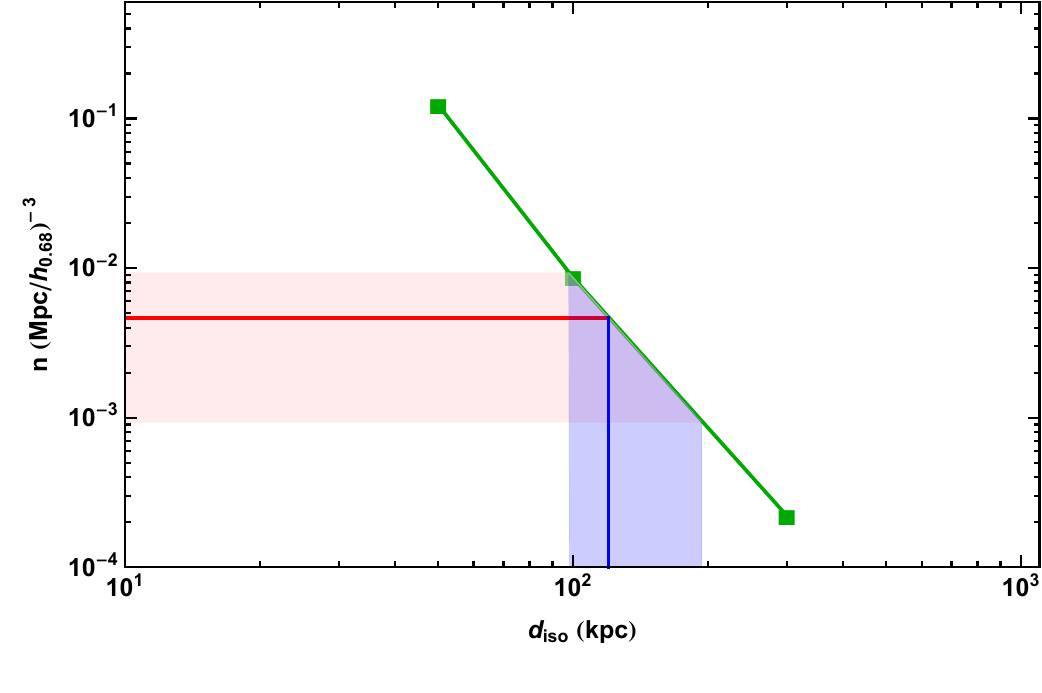}
\caption{
Asymptotic ($z=10$) number density of SMBHs, $n$, versus $d_{\rm
  iso}$. The green squares show results for the analysis with $d_{\rm
  iso}=30$, 50 and 100~kpc, joined by the green lines assuming a power
law dependence of $n$ on $d_{\rm iso}$. The red line indicates
$n_{\rm{SMBH}}(z=0)$ corresponding to $L_{\rm min}=0.33L_*$ and the
red band shows the range of $n_{\rm{SMBH}}$ corresponding to $L_{\rm
  min}=0.1$ to 1~$L_*$. The blue line and band show the corresponding
value and range of $d_{\rm iso}$ implied by this range of $n$.}
\label{fig:ndenvsdiso}
\end{figure*}

Note when comparing with $n_{\rm SMBH}(z=0)$ that the model $n$ does
not account for any decrease due to merging of SMBHs. However, we can
assess how many mergers occur in the simulation: for $d_{\rm
  iso}=50$~kpc only $\sim0.2$\% of SMBHs suffer a merger with another
SMBH by $z=10$, i.e., it is a very minor effect. Some additional SMBH
mergers will occur at $z<10$, but given the low rate of merging at
$z>10$ and the low fraction of $>10^9\:M_\odot$ halos at $z=10$ that
host SMBHs (for $d_{\rm iso}=50$~kpc this is $\simeq0.16$, discussed
below), it seems unlikely that this will lead to more than a factor of
three reduction in $n$. Thus we consider that the cases of $d_{\rm
  iso}=50$ and 100~kpc are the most relevant in the context of this
model of Pop III.1 seeds for forming the whole cosmic population of
SMBHs. From here on we will focus on the cases between $d_{\rm
  iso}=30$ and 300~kpc.

If we assume SMBH mergers are negligible, then we can use the
asymptotic (i.e., $z=10$) number density of SMBH remnants for the
cases of $d_{\rm iso}=50$ to 300~kpc to estimate the range of $d_{\rm
  iso}$ that is implied by our adopted constraint on $n_{\rm
  SMBH}(z=0)$. This is shown in Figure~\ref{fig:ndenvsdiso} as a blue
shaded band, i.e., implying $100\:{\rm kpc}\lesssim d_{\rm iso}
\lesssim 200$~kpc.

We have also checked the sensitivity of our results to the choice of
$1\times10^6\:M_\odot$ as the threshold halo mass for leading to a Pop
III.1 source. As discussed above in \S\ref{S:intro}, effects such as
dark matter particle streaming velocities relative to baryons may
increase this threshold mass (e.g., Fialkov et al. 2012). However, we
find that raising the threshold mass by a factor of four, i.e., to
$4\times10^6\:M_\odot$, has a very minor ($\lesssim20\%$) effect on
the overall number density of the sources at late times, which is much
smaller than the variation resulting from the choice of $d_{\rm iso}$.

\begin{figure*}
\centering
\includegraphics[scale=0.65]{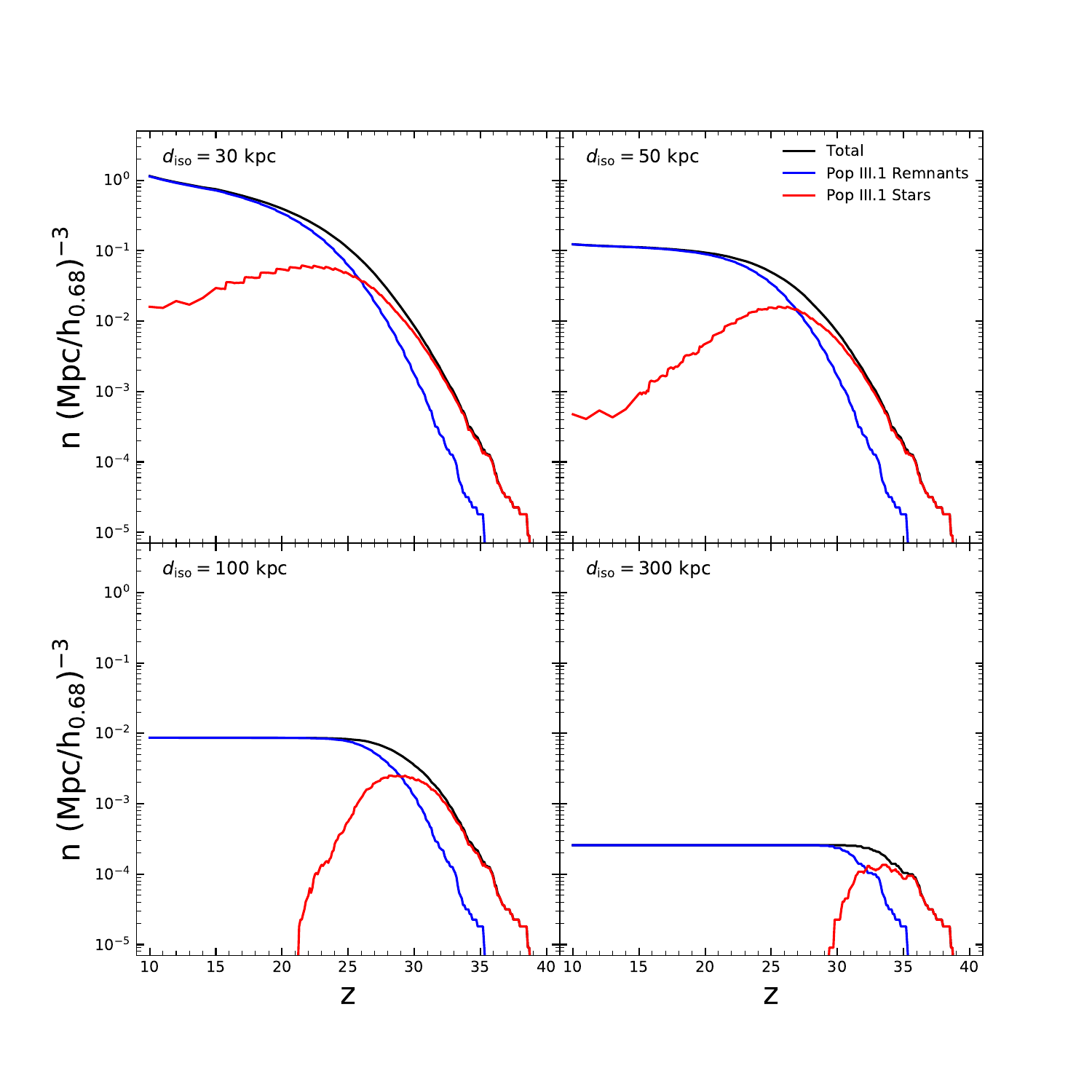}
\caption{
Evolution of the comoving number density of Pop III.1 stars (red
lines), SMBH remnants (blue lines) and the total number of sources
(black lines) for $d_{\rm iso}=$30~kpc (top left), 50~kpc (top right),
100~kpc (bottom left) and 300~kpc (bottom right). All models shown here assume a Pop III.1 Star formation time and/or lifetime, $t_{*f}$, of 10~Myr.
}
\label{fig:nevol_stars}
\end{figure*} 

\begin{figure*}
\centering
\includegraphics[scale=0.65]{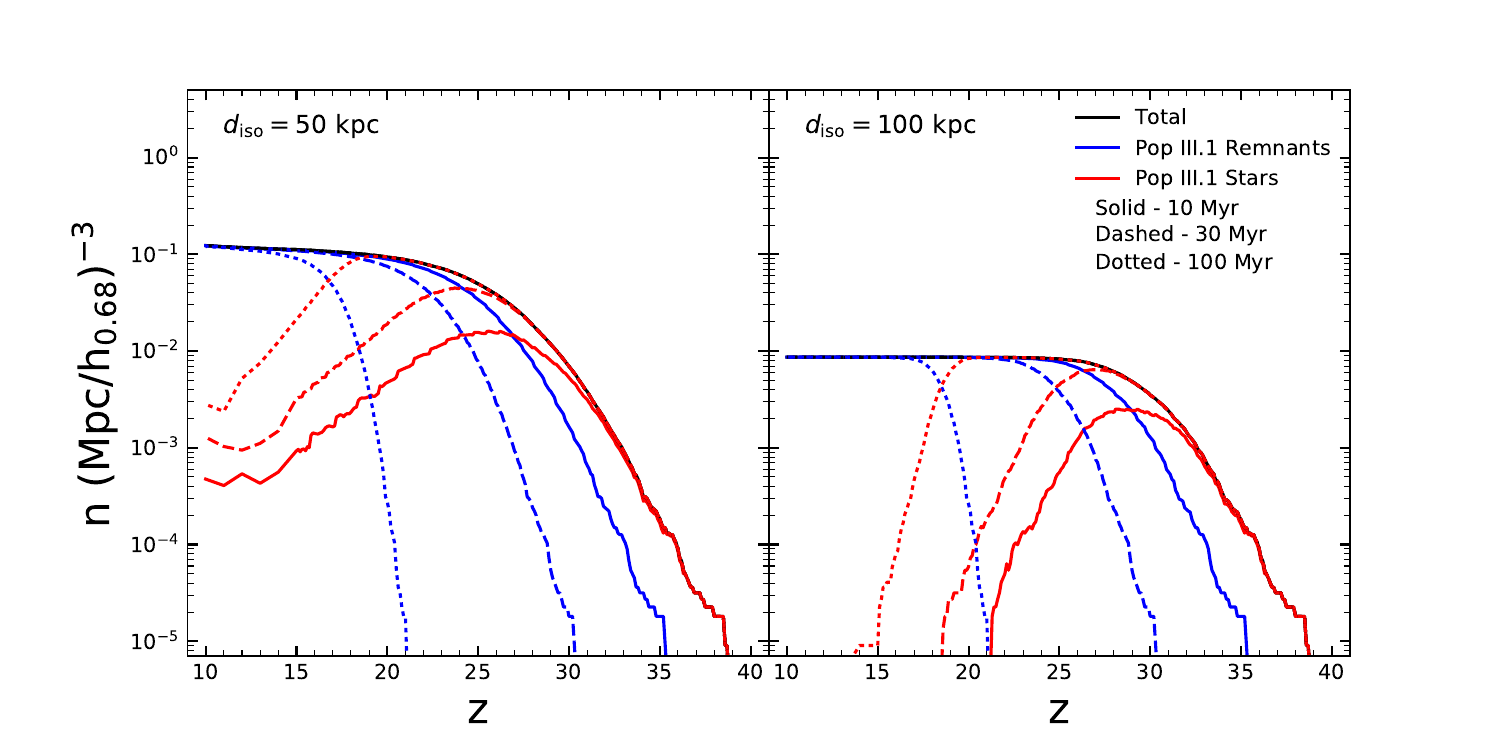}
\caption{
Effect of varying Pop III.1 star formation time and/or lifetime,
$t_{*f}$ on the evolution of the comoving number density of Pop III.1
stars (red lines), SMBH remnants (blue lines) and the total number of
sources (black lines) for $d_{\rm iso}=$50~kpc (left) and 100~kpc
(right). Within each panel, different line styles of solid, dashed and
dotted represent cases with $t_{*f}=10, 30, 100$~Myr, respectively. }
\label{fig:nevol_stars2}
\end{figure*} 

Figure~\ref{fig:nevol_stars} shows the separate evolution of the
comoving number density of Pop III.1 stars, SMBH remnants and the
total of these two components for $d_{\rm iso}=30$, 50, 100 and
300~kpc for the case with $t_{*f}=10$~Myr. Pop III.1 stars dominate at
very early times, while SMBHs dominate at later times. For example,
for $d_{\rm iso}\sim100$~kpc the Pop III.1 star formation rate (SFR)
peaks at $z\simeq30$ and effectively stops below $z\simeq 25$, while
for smaller values of $d_{\rm iso}$ there can still be significant new
Pop III.1 sources forming or existing at $z\sim 10$ to 15. We have
seen that from the overall number of SMBH remnants produced in
comparison to observed local comoving number densities of SMBHs that
the models with $d_{\rm iso} = 50$ and 100~kpc are the most relevant
if all SMBHs are to form via Pop III.1 seeds. Thus in
Figure~\ref{fig:nevol_stars2} we focus on these two cases and explore
the effect of varying the overall time that Pop III.1 stars exist
(i.e., combining their formation and subsequent lifetime before
collapsing to SMBHs), $t_{*f}$, with values explored of 10, 30 and
100~Myr. Extending the duration of the Pop III.1 Star phase causes
them to be present down to lower redshifts, but, in the context of our
modeling, does not affect the final number density of the SMBH
remnants. For $d_{\rm iso}=50$~kpc and $t_{*f}=100$~Myr, significant
number densities of Pop III.1 stars can be present down to z=10, but
still at levels that are about a factor of 30 smaller than at the peak
at $z\sim20$. Variation in $t_{*f}$ also affects the redshift when the
first SMBHs, i.e., AGN, appear. For $t_{*f}=10$~Myr, SMBHs start
appearing at $z\sim35$ and the populations are largely in place by
$z\sim25$. However, for $t_{*f}=100$~Myr, SMBHs do not appear until
$z\sim20$. Thus the detection or non-detection of Pop III.1
supermassive stars and/or accreting SMBHs at $z\simeq10$ to 15,
potentially possible with the {\it James Webb Space Telescope} ({\it
  JWST}) (Freese et al. 2010), could help to distinguish between these
models.

\subsection{Effects of Cosmic Variance and $\sigma_8$}\label{S:variance}

\begin{figure*}
 \centering
 \includegraphics[scale=0.6]{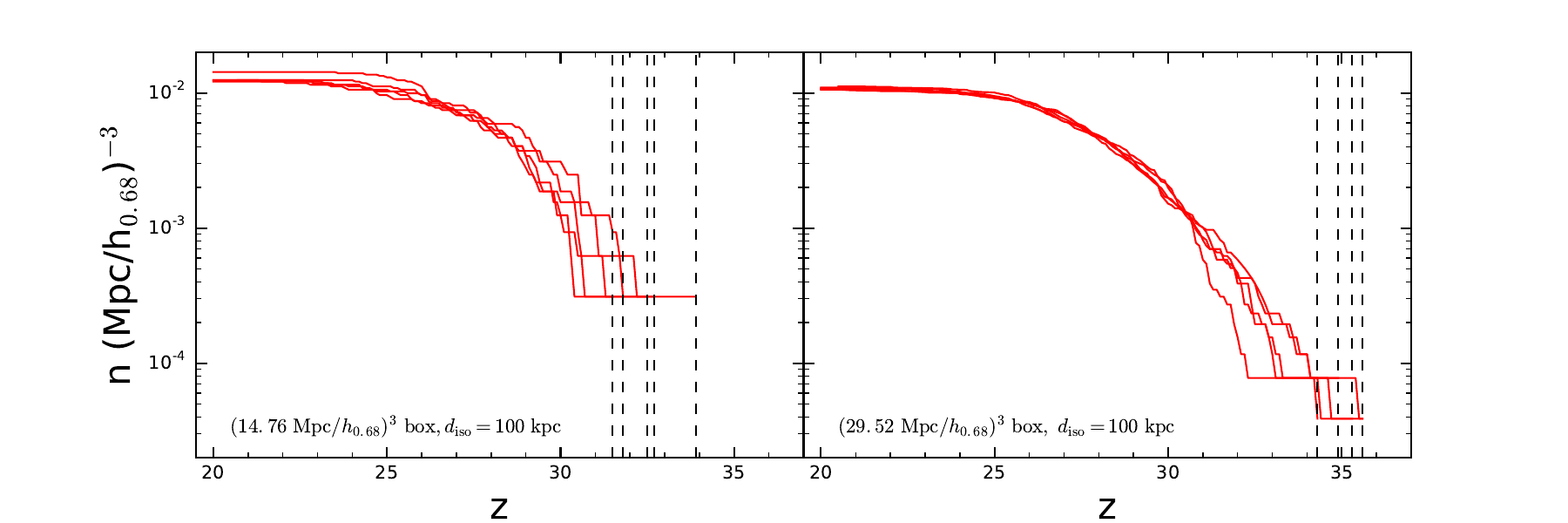}
 \caption{ 
{\it (a) Left:} Evolution of the comoving total number density of Pop
III.1 stars and their SMBH remnants for $d_{\rm iso}=100$~kpc in five
realizations of a $(14.76 h_{0.68}^{-1}\:{\rm Mpc})^3$ volume. The
vertical dashed lines indicate the redshift of the appearance of the
first halo in each run.
{\it (b) Right:} As (a), but now showing results for five realizations
of a $(29.52 h_{0.68}^{-1}\:{\rm Mpc})^3$ volume.
}
\label{fig:variance}
\end{figure*}

We study how the results are affected by cosmic variance by running
several simulations of smaller volumes of $(10~ h^{-1}~ {\rm Mpc})^3 =
(14.76\ h_{0.68}^{-1}~ {\rm Mpc})^3 $ and $(20\ h^{-1}~ {\rm Mpc})^3 =
(29.52\ h_{0.68}^{-1}~ {\rm Mpc})^3$. For each of these volumes, five
independent simulations were run using different random seeds to
generate the initial conditions. Figure~\ref{fig:variance} shows the
results of these runs for the number density evolution of Pop III.1
stars and SMBH remnants for the case of $d_{\rm iso}=100$~kpc. We see
that while there is moderate variation in redshift of the first Pop
III.1 source in each volume, the dispersion in the final number
densities of sources (i.e., for $z\lesssim25$) in these simulations is
very minor.

\begin{figure}
\centering
\includegraphics[width=\linewidth]{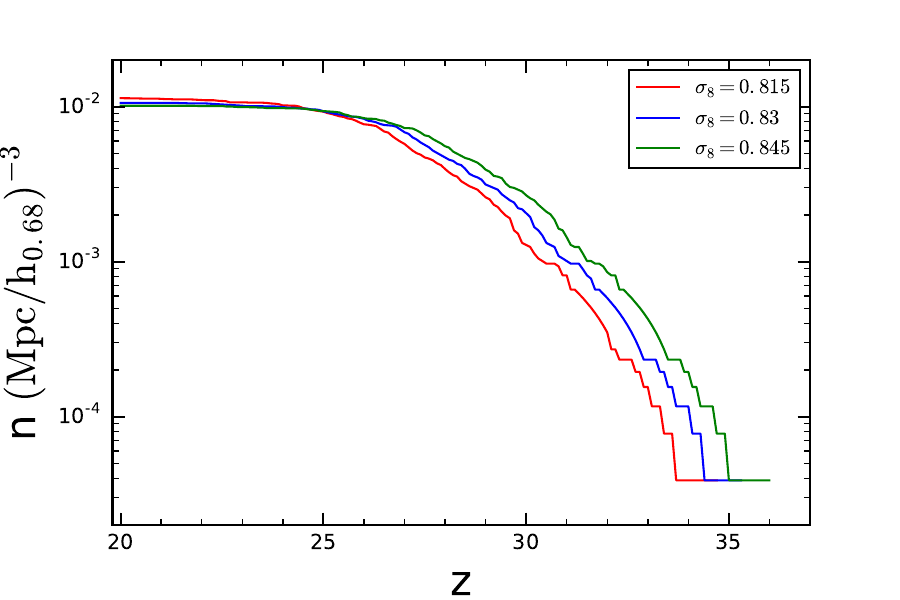}
\caption{
Effect of $\sigma_{8}$ uncertainties on the evolution of comoving
total number density of Pop III.1 stars and their SMBH remnants for
$d_{\rm iso}=100$~kpc, based on simulations of a
$(29.52\ h_{0.68}^{-1}\:{\rm Mpc})^3$ volume.}
\label{fig:sigma8}
\end{figure}

Halo number densities will depend on cosmological parameters.  In
particular, the number of rare objects is mostly sensitive to the
normalization of the power spectrum via $\sigma_8$. Thus, next we
examine how the choice of $\sigma_8$ affects the number density of Pop
III.1 stars and SMBHs. We explore a range of $\sigma_8=0.830 \pm0.015$
(Planck collaboration 2015). Figure~\ref{fig:sigma8} shows the effect
of varying $\sigma_8$ by this amount on the total number density of
Pop III.1 stars and SMBHs for the case of $d_{\rm iso}=100$~kpc for a
simulation of a $(29.52\ h_{0.68}^{-1}~ {\rm Mpc})^3$ volume. Again
the dispersion in the final number densities of sources (i.e., for
$z\lesssim25$) in these simulations is very minor, i.e.,
$\lesssim7\%$.

\subsection{Mass Function of SMBH-Host Halos}\label{S:massfn}

\begin{figure*}
\centering
\includegraphics[scale=0.7]{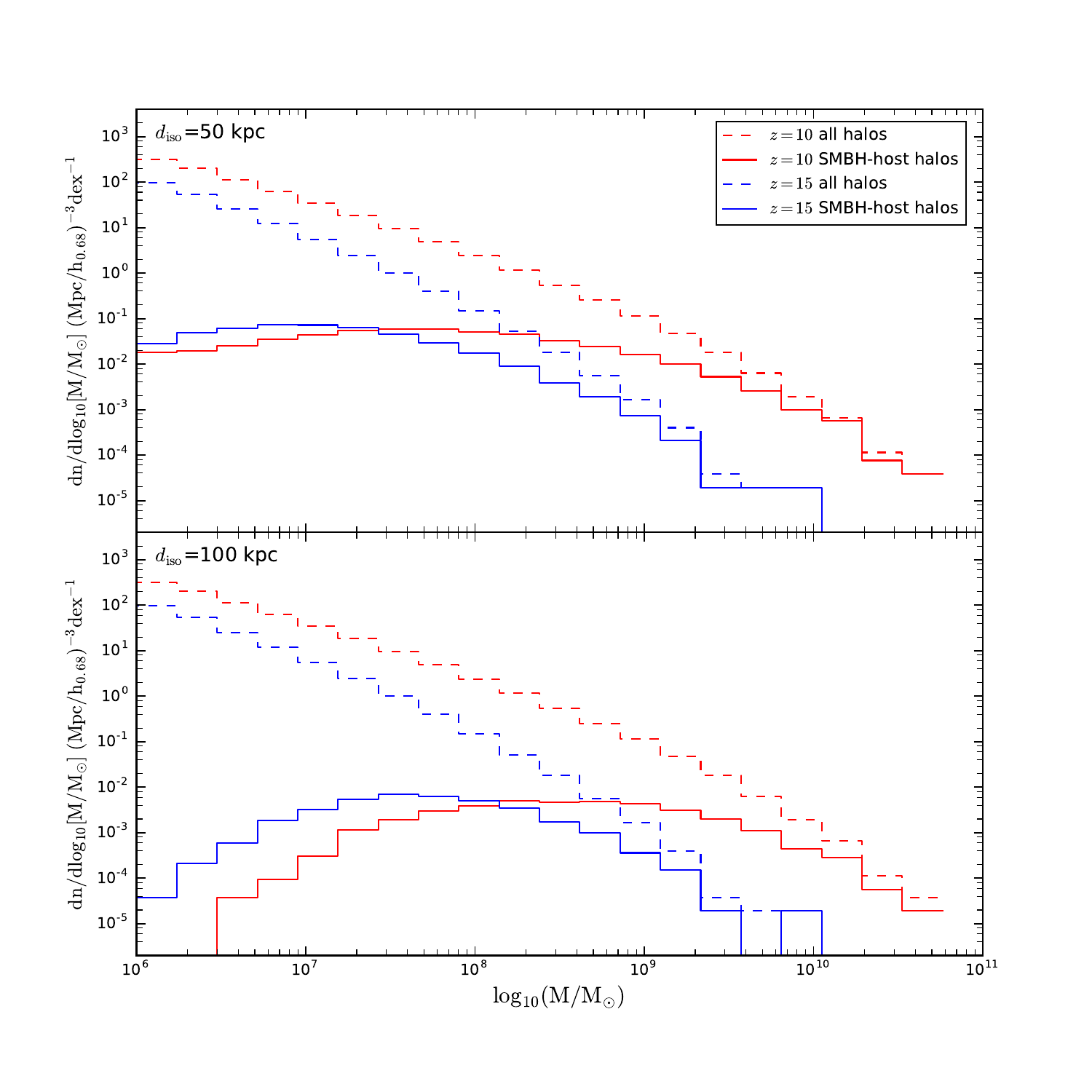}
\caption{
The mass functions of SMBH-host halos (solid lines) compared to all
halos (dashed lines) at $z=10$ (red lines) and $z=15$ (blue lines).
Results for $d_{\rm iso}=50$~kpc are shown in the top panel, while
those for $d_{\rm iso}=100$~kpc are shown in the bottom panel. These
results are based on the fiducial simulation of a
$(60.47\ h_{0.68}^{-1}~ {\rm Mpc})^3$ volume.}
\label{fig:mass}
\end{figure*}

The halos that form Pop III.1 stars and host their SMBH remnants are
then followed to lower redshifts, as far as $z=10$. These halos grow
in mass by accreting dark matter particles and merging with other
identified halos. As described in \S\ref{S:methods}, in a merger the
more massive halo retains its identity. For values of $d_{\rm
  iso}\gtrsim 50$~kpc, the halos that are merging with the
SMBH-hosting halos are typically of lower mass and do not host
SMBHs. Occasionally, they are more massive, in which case the
SMBH-hosting character of the halo is transferred to this new halo
identity. Even more rare is a merger of two halos that both host
SMBHs. The properties of these binary SMBH halos and predictions for
the eventual merger of the SMBHs will be presented in a future paper
in this series. Here we focus simply on the mass function of the
SMBH-hosting halos: Figure~\ref{fig:mass} shows the distribution of
these masses at $z=10$ and 15 for cases of $d_{\rm iso}=50$ and
100~kpc. For comparison, we also show the mass function of all halos
at $z=15$ and 10 with dashed histograms.

For $d_{\rm iso}=100$~kpc we see that at $z=15$ the typical SMBH-host
halo has $\gtrsim 10^7\:M_\odot$, while by $z=10$ this grows to
$\gtrsim 10^8\:M_\odot$ (although note there can be some SMBHs in
lower mass halos). In comparison, for $d_{\rm iso}=50$~kpc, SMBHs are
more numerous and at a given redshift tend to occupy lower mass halos.

In general, the most massive halos, $>10^{10}\:M_\odot$, have
the highest occupation fractions of SMBHs. By $z=10$ this occupation
fraction is close to unity for $d_{\rm iso}=50$~kpc and slightly
smaller for $d_{\rm iso}=100$~kpc. However, the fraction of
$>10^9\:M_\odot$ halos at $z=10$ that host SMBHs is only 0.16 for
$d_{\rm iso}=50$~kpc and 0.05 for $d_{\rm iso}=100$~kpc. These small
fractions indicate the sparseness of SMBHs among these early galaxies
for these relatively large values of $d_{\rm iso}$. This suggests that
in these models mergers of SMBHs will continue to be relatively rare
at $z<10$, especially for the $d_{\rm iso}=100$~kpc case.

\subsection{Synthetic Sky Maps}\label{S:maps}

Given the considerations of the local SMBH number density and the
results shown in Figure~\ref{fig:nevol}, we again focus on the cases
of $d_{\rm iso}=50$ and 100~kpc, with the latter being the preferred,
fiducial case. For reference, at $z=10, 15, 20, 30, 40$ the angular
size of the box is $21.53\arcmin$, $19.84\arcmin$, $18.93\arcmin$,
$17.93\arcmin$, $17.37\arcmin$, respectively.

\begin{figure*}
\centering
\includegraphics[scale=1.]{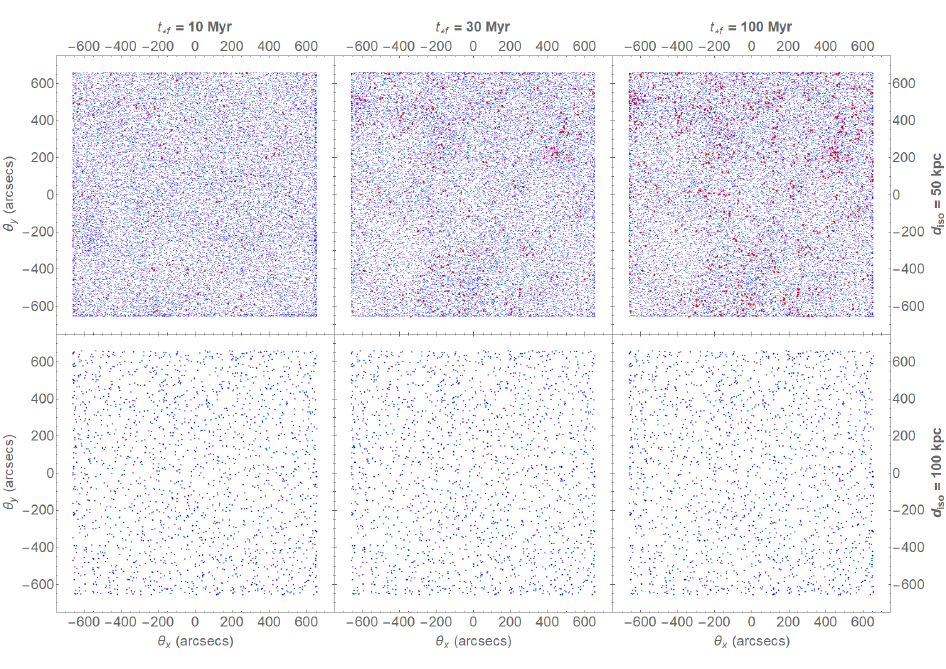}

\caption{
Synthetic sky maps at $z=10$ of the $(60.47~h_{0.68}^{-1}~\rm{Mpc})^3$
comoving box (projection equivalent to $\Delta z=0.28$) of the $d_{\rm
  iso}=50$~kpc model (top row) with Pop III.1 stars (red dots) and
SMBH remnants (blue dots). The panels from left to right show the
cases with $t_{*f}=10, 30, 100$~Myr. The bottom row shows the same for
$d_{\rm iso} = 100$~kpc (but note that in these cases there are no Pop
III.1 stars left by $z=10$ even for $t_{*f}=100$~Myr).}\label{fig:maps}
\end{figure*}  

To make an approximate synthetic sky map of the SMBH population (which
may manifest themselves as AGN), we simply project through the entire
volume of the box, adopting a constant, fixed redshift. This
approximation ignores the finite light travel time across the
thickness of the box, which means that the projection of the total
source population is roughly equivalent to observing a finite redshift
interval of the real universe. At $z=10$, 15, 20 these redshift
intervals are $\Delta z=$0.28, 0.49, 0.75, respectively. Of course for
direct comparison with AGN populations one would also need to model
the duty cycle of emission and the luminosity function and spectral
energy distribution properties of the accreting SMBHs. Such modeling
requires making many uncertain assumptions and is beyond the scope of
the present paper, but will be addressed in future studies. More
direct comparisons of the sky maps of the sources can in principle be
done with other theoretical models and simulations that also aim to
predict SMBH locations, along with the angular correlation function of
the sources, discussed below.

Figure~\ref{fig:maps} shows maps of the Pop III.1 Star and SMBH
remnant populations for $d_{\rm iso}=50, 100, 300$~kpc at $z=10$, for
the three different values of $t_{*f}=10, 30, 100$~Myr. For smaller
$d_{\rm iso}$ and longer $t_{*f}$, there are greater numbers of Pop
III.1 stars present at $z=10$.

This figure shows the dramatic effect of $d_{\rm iso}$ on the number
of SMBH remnants predicted by the model, i.e., there is about a factor
of 10 reduction in the number density of SMBHs on increasing $d_{\rm
  iso}$ from 50~kpc to 100~kpc. The angular clustering properties of
these sources will be examined below.

\begin{figure*}
\centering
\includegraphics[width=\linewidth]{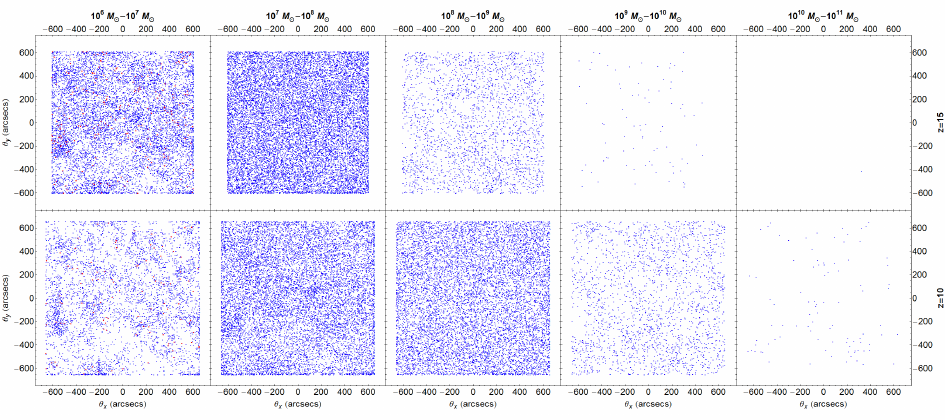}
\includegraphics[width=\linewidth]{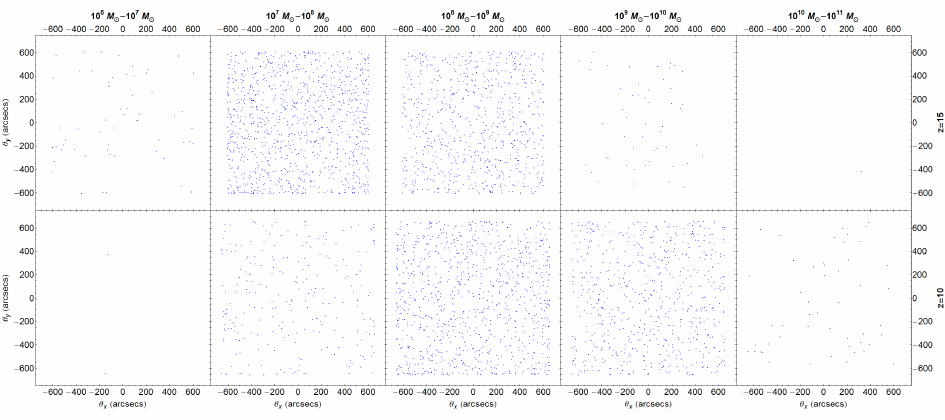}
\caption{
{\it (a) Top panel:}
Synthetic maps at $z=15$ (top row) and $z=10$ (bottom row) of $d_{\rm
  iso}=50$~kpc Pop III.1 stars (assuming $t_{*f}=10\:$Myr) (red dots)
and SMBH remnants (blue dots) for mass bins from left column to right
column of $10^{6}-10^{7}\:M_{\odot}$, $10^{7}-10^{8}\:M_{\odot}$,
$10^{8}-10^{9}\:M_{\odot}$, $10^{9}-10^{10}\:M_{\odot}$ and
$10^{10}-10^{11}\:M_{\odot}$.
{\it (b) Bottom panel:} As (a), but now for $d_{\rm iso} = 100$~kpc. Pop III.1 stars are present at $z=10$ and $z=15$ only for $d_{\rm iso}=50$~kpc and they have mass in the range $10^{6}-10^{7}\:M_{\odot}$. } 

\label{fig:maps_mass}
\end{figure*}

Figure~\ref{fig:maps_mass} shows synthetic maps at $z=10$ and 15, but
now separating out different mass halos that are hosting Pop III.1
stars and SMBHs for the cases of $d_{\rm iso}=50$ and 100~kpc (for
$t_{*f}=10$~Myr). The evolution of the typical SMBH host halo towards
higher masses as the universe evolves towards lower redshift can be
seen. The lower mass halos tend to be the SMBHs that have formed most
recently and, at least in the $d_{\rm iso}=50$~kpc case where there
are significant numbers, these show distinctive clustering properties
compared to the more massive, typically older, sources.

\subsection{Angular Correlation}\label{S:angular}

We calculate the two point angular correlation function (2PACF) of
$d_{\rm iso} = 50$ and $100\: \rm{kpc}$ SMBH remnants. The angular
correlation function tells us how the projection of these remnants is
correlated compared to a Poisson distribution. We use a random catalog
with 50 times the number of halos as in the SMBH sample, implementing
the Landy-Szalay estimator to calculate the angular correlation
function, $\omega(\theta)$:
\begin{equation}
\omega(\theta)=\frac{DD-2DR+RR}{RR},
\end{equation}
where $DD$ represents the total weight of pairs of halos from the data
(i.e., the SMBHs) within each bin, $RR$ represents the weight of pairs
from the random Poisson catalog, and $DR$ represents the weight of
pairs with one particle from the data and one from the random
catalog. We used the \textit{TreeCorr} code (Jarvis et al. 2004) for
these angular correlation calculations.
 
\begin{figure*}
\centering
\includegraphics[scale=0.7]{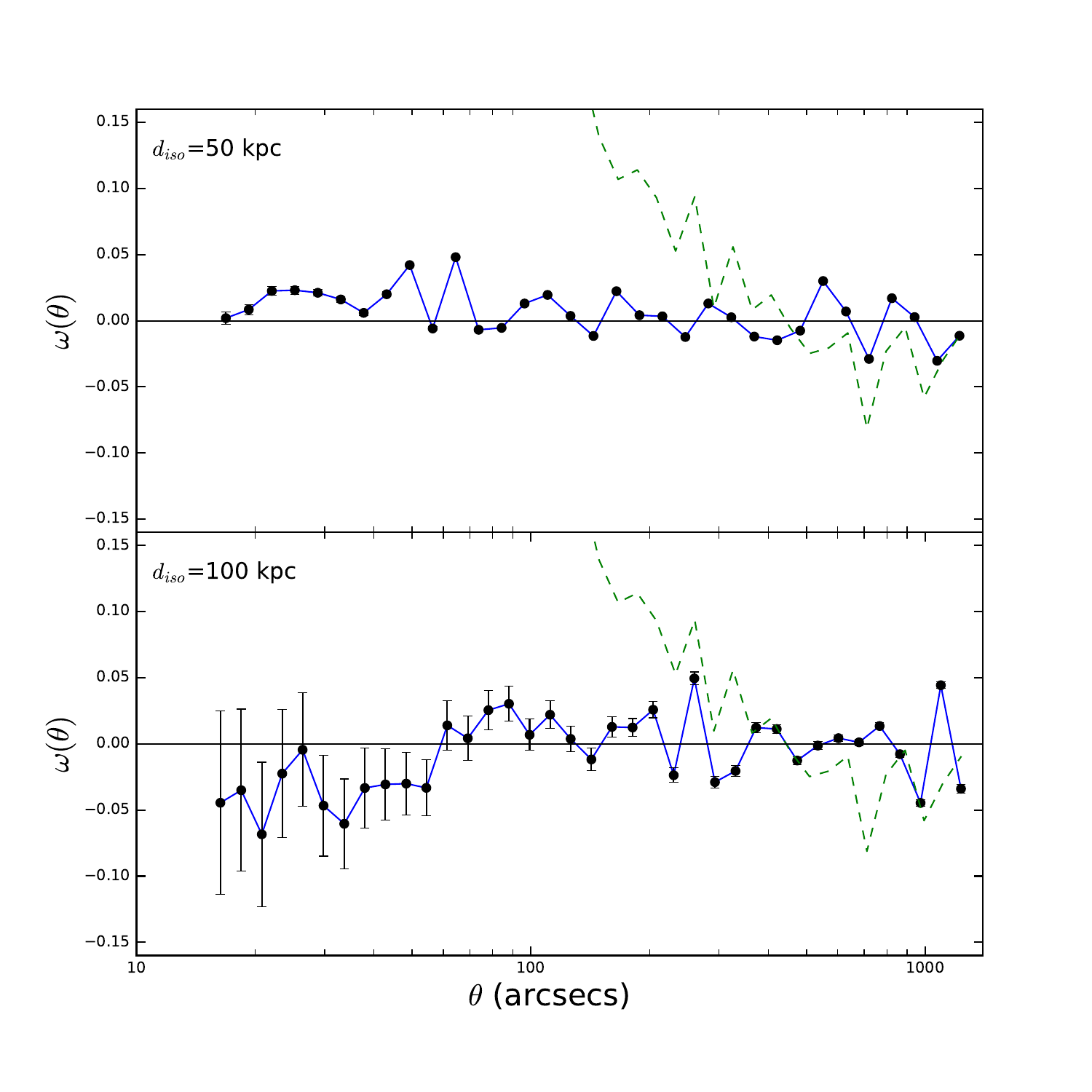}

\caption{
2PACF (black points and blue solid line) of $d_{\rm iso}=50\:$kpc (top
panel) and $d_{\rm iso}=100\:\rm{kpc}$ (bottom panel) remnants at
$z=10$. The error bars indicate shot noise in each angular separation
bin. For comparison we have also plotted the 2PACF of all the halos
with masses $>10^9\:M_{\odot}$ at $z=10$ (green dashed lines). It
rises sharply at small scales, which shows they are highly clustered
(it reaches a value of $0.22$ at 100\arcsec\ and $0.57$ at
20\arcsec). The SMBHs in these models show much lower levels of
clustering on small angular scales due to feedback suppression of
neighboring halos that prevents them from being Pop III.1 sources.}
\label{fig:angcorr}
\end{figure*}

Figure~\ref{fig:angcorr} shows $\omega(\theta)$ for the cases of
$d_{\rm iso}=50$~and 100~kpc observed at $z=10$ (i.e., derived from
the spatial distributions shown in Fig.~\ref{fig:maps} for the case
with $t_{*f}=10$~Myr). These have 27,122 and 1,913 SMBH sources,
respectively.  To increase the signal to noise we have observed the
simulation volume at $z=10$ from the three orthogonal directions and
combined the results. Overall the 2PACFs of both cases are relatively
flat, especially compared to that of all halos with masses
$>10^9\:M_\odot$, which is shown by the green dashed line in these
figures. For $d_{\rm iso}=50$~kpc, there is a sign of modest excess
clustering signal on angular scales $\lesssim 50\arcsec$. For $d_{\rm
  iso}=100$~kpc, with a smaller number of sources and thus larger
Poisson uncertainties, there is hint of a decrease of $\omega(\theta)$
below the Poisson level on scales $\lesssim50\arcsec$. This could be
related to the angular scale of the 100~kpc proper distance of the
isolation (i.e., feedback) distance $d_{\rm iso}$ at $z\sim30$, which
corresponds to a comoving distance of $\sim3$~Mpc.  By $z=10$ this
corresponds to an angular scale of 64\arcsec. Thus the signature of
feedback cleared bubbles, which have a deficit of SMBHs due to
destruction of Pop III.1 seeds, may be revealed in the angular
correlation function. In particular, effective feedback suppression of
neighboring sources leads to a relatively flat angular correlation
function.

\section{Discussion and Conclusions}\label{S:disc}

We have presented a simple model for the formation of supermassive
black holes from the remnants of special Population III stars, i.e.,
Pop III.1 stars that form in isolation from other astrophysical
sources. The physical mechanism motivating this scenario involves the
Pop III.1 protostar being supported by WIMP dark matter annihilation
heating (Spolyar et al. 2008; Natarajan et al. 2009; Rindler-Daller et
al. 2015), so that it retains a large photospheric radius while it is
accreting. This reduces the influence of ionizing feedback on the
protostar's own accretion (McKee \& Tan 2008; Hosokawa et al. 2011),
allowing it gather most of the $\sim10^5\:M_\odot$ of baryons present
in the host minihalo. Such a mechanism may naturally explain the
  dearth of SMBHs with masses below $\sim10^5\:M_\odot$. While many
of the details of this scenario remain to be explored, our approach
here has been to focus on the Pop III.1 star and remnant SMBH
populations that are predicted to form in such a model, particularly
their dependencies on the isolation distance, $d_{\rm iso}$, needed
for Pop III.1 star formation.

Assuming that all SMBHs are produced by this mechanism, we have found
that to produce the required number density of sources that can
explain local ($z=0$) SMBH populations requires $d_{\rm iso}\lesssim
100$~kpc (proper distance). For the model with $d_{\rm iso}=100$~kpc
and SMBH formation times $t_{*f}=10$~Myr, the formation of Pop III.1
stars and thus SMBHs starts to become significant at $z=35$, peaks at
$z\simeq 30$ and is largely complete by $z=25$, i.e., occurring in a
period from only $\sim$80~Myr to $\sim$130~Myr after the Big
Bang. This result can be understood with a simple physical model: the
comoving number density of SMBHs is $n_{\rm SMBH} \sim (4\pi
\tilde{d}_{\rm iso,z=30}^3/3)^{-1}\rightarrow 8.8\times10^{-3}
(\tilde{d}_{\rm iso,z=30}/3\:{\rm Mpc})^{-3}\:{\rm Mpc}^{-3}$, where
$\tilde{d}_{\rm iso,z=30}$ is the comoving distance at $z=30$
corresponding to proper distance $d_{\rm iso}$. Note, with such values
of $d_{\rm iso}$ we do not expect SMBH mergers to be too significant
in reducing their global comoving number density. Full predictions of
merger rates evaluated down to $z=0$ will be presented in a future
paper.

Compared to the ``direct collapse'' scenario of SMBH formation (e.g.,
Chon et al. 2016), the Pop III.1 protostar progenitor model we have
presented involves much earlier and widespread formation of
SMBHs. Thus there are expected to be significant differences in AGN
luminosity functions at $z\gtrsim20$ between these models, i.e., there
are much larger number densities of AGNs at these high redshifts in
the Pop III.1 formation scenario.

We have followed the SMBH population to $z=10$. By this redshift,
SMBHs tend to reside in halos $\gtrsim 10^8\:M_\odot$, extending up to
$\sim10^{11}\:M_\odot$. The angular correlation function of these
sources at $z=10$ is very flat, with little deviation from a random
distribution, which we expect is a result of a competition between
source feedback and the intrinsic clustering expected from
hierarchical structure formation.

The models presented here, being first approaches for describing the
cosmic distribution of Pop III.1 sources, are intended to be simple
and involve relatively few free parameters. Of course, much more
detailed exploration of the growth and feedback of supermassive Pop
III.1 stars and their accreting SMBH remnants is still needed in the
context of this scenario, both to explore the viability of forming
SMBHs from Pop III.1 sources themselves and how their local feedback
may limit other Pop III.1 star and SMBH formation, i.e., setting the
value of $d_{\rm iso}$. Diffuse feedback, e.g., from a Lyman-Werner
FUV background due to widespread early stellar populations, i.e., Pop
III.2 and Pop II stars, may also need to be considered depending on
the formation efficiencies and IMFs of these populations. Such a
diffuse background feedback could act to effectively truncate new Pop
III.1 star formation below a certain redshift, independent of local
feedback that we have so far parameterized via $d_{\rm iso}$.

We defer exploration of these types of models, which have additional
free parameters, to future studies. Other future work will include
tests of the models that involve modeling the potentially observable
luminosity functions of Pop III.1 and SMBH sources at high redshift
and following the populations of SMBHs down to $z=0$ to compare with
the observed clustering properties of the local SMBH population.

{\textbf{Acknowledgments} --} We thank Peter Behroozi, Krittapas
Chanchaiworawit, Mark Dijkstra, Alister Graham, Lucio Mayer, Rowan
Smith, Kei Tanaka, Brian Yanny and an anonymous referee for helpful
comments and discussions. Fermilab is operated by Fermi Research
Alliance, LLC, under Contract No. DE-AC02-07CH11359 with the
U.S. Department of Energy. NB acknowledges the support by the Fermilab
Graduate Student Research Program in Theoretical Physics. NB also
acknowledges the support of the D-ITP consortium, a programme of the
Netherlands Organization for Scientific Research (NWO) that is funded
by the Dutch Ministry of Education, Culture and Science (OCW). JCT
acknowledges support from NSF grant AST 1212089 and ERC Advanced Grant
78882 (MSTAR).


\end{document}